\documentclass[conference]{IEEEtran}
\IEEEoverridecommandlockouts
\usepackage{cite}
\usepackage{amsmath,amssymb,amsfonts}
\usepackage{algorithmic}
\usepackage{graphicx}
\usepackage{textcomp}
\usepackage{xcolor}
\usepackage{url}
\usepackage{hyperref}
\def\BibTeX{{\rm B\kern-.05em{\sc i\kern-.025em b}\kern-.08em
    T\kern-.1667em\lower.7ex\hbox{E}\kern-.125emX}}

\begin{document}

\title{Energy-Based Fair Queuing Scheduler for \\ Mobile Systems\\
}

\author{\IEEEauthorblockN{Deshpande Mandar Anil}
\IEEEauthorblockA{\textit{Department of Computer Science and Engineering} \\
\textit{National Institute of Technology, Tiruchirappalli}\\
Tiruchirappalli - 620015, Tamil Nadu, India \\
deshpande.mandar.a@gmail.com}
}

\maketitle

\begin{abstract}
This paper presents a concrete implementation and a comprehensive assessment of the Energy-based Fair Queuing (EFQ) scheduling algorithm based on the Linux operating system. EFQ is an extended application of the classical fair queuing algorithm in the energy domain. It is designed to provide proportional power sharing as well as effective time-constraint compliance in energy-centric Power Management (PM) schemes, a type of operating system-level PM scheme targeted at providing a battery lifetime guarantee for energy-limited mobile systems. In this work, the structure of the Linux Completely Fair Scheduler (CFS) has been effectively utilized to ease the EFQ implementation and to reduce the scheduling overhead. To assess the properties of the EFQ scheduler, a test-bench based on various types of load has been developed and tested. The EFQ algorithm is assessed from two aspects: energy management and real-time scheduling. Experimental results on energy management show that EFQ is more effective than the CFS scheduler in managing energy and it can achieve a proportional share of the system power regardless of the device on which the energy is spent. Experimental results on real-time scheduling demonstrate that EFQ can achieve strict time-constraint compliance and a robust response time when the energy estimation error and the number of active tasks increase.
\end{abstract}


\section{Introduction}
Currently, Operating System (OS)-based mobile devices are experiencing significant growth in performance and functionality, making them capable of handling multiple applications simultaneously to meet the diversity of user needs. The increased complexity and functionality in many Consumer Electronic (CE) devices has motivated a transformation of system usefulness assessment from a Quality of Service (QoS) approach to a Quality of Experience (QoE) approach \cite{b1,b2}. In the meantime, due to the increasing energy demand of applications and the relentless trend to make the battery lighter and smaller in CE devices, energy in modern mobile systems has become a limited resource as important as the CPU, memory, and network \cite{b5,b6,b7}.

When a user is running multiple applications with different preferences on a battery-limited mobile device, such as a laptop or a tablet, they usually have a requirement concerning how long the battery needs to last for the most-preferred applications. The expected lifetime of the system is an important factor in the total user experience \cite{b2,b3}. A failure to achieve the lifetime will significantly degrade the QoE and make it unacceptable to the user.

To guarantee a user-specified battery lifetime for a mobile system, it has been proposed that energy, instead of CPU or transmission bandwidth, should be managed as the first-class resource in the system \cite{b6,b7,b8}. Managing the energy as the first-class resource allows the limited energy to be properly distributed both over time and among the applications. A proper energy distribution over time sets the average system power required to achieve a target lifetime, while a proper energy distribution among different applications guarantees an energy allocation to the most-preferred application that is consistent with its actual energy demand. Specifically, when there is not enough energy available to maintain all applications running normally, the energy allocation of those least-preferred applications should be properly restricted so that enough remaining energy can be reserved for the most-preferred applications to run them normally until the expected lifetime is reached.

Achieving a user-desired lifetime is only the first step to guaranteeing an acceptable user experience. With a proper energy allocation reserved for a time-sensitive task, the scheduler needs to work intelligently to let the task consume its share of energy within the time-constraint. A better user experience is achieved when the performance of those most-preferred applications is always guaranteed during the lifetime. Moreover, if any highly preferred application becomes ill-behaved by demanding excessive energy in a short time, the scheduler should maintain the performance of other preferred applications by restricting any power pulse caused by the abnormal application.

Power Management (PM) schemes from the operating system level are a promising method to manage the battery lifetime because the OS is aware of the battery discharging status, the application energy requests, as well as the device power states. However, most of the OS-level PM schemes only partially utilize the energy-related information that is available to them. They are usually device-specific and performance-centric; energy saving is only focused on specific devices and is made with best effort under minimum performance constraints. Such a scheme is too weak to ensure a user-specified battery lifetime. To fully utilize the energy-related information in the OS and to develop strong energy-aware PM schemes that are able to provide battery lifetime guarantees, energy should be globally managed as the first-class resource in the system, and the corresponding PM schemes are called energy-centric PM schemes. An energy-centric PM scheme can ensure a target battery lifetime for user-specified applications by restricting the battery discharging rate over time and proportionally allocating energy among different applications. It relies on the cooperation of the energy estimation module, energy accounting module, energy allocation module, and energy scheduling module to achieve advanced energy objectives such as a lifetime guarantee. Many research works have been done in the domains of energy estimation \cite{b13,b14}, energy accounting \cite{b14,b16,b17}, and energy allocation \cite{b14,b16}; however, little attention has been paid to energy-centric scheduling, which is pivotal in achieving proportional energy use and ensuring application performance.

The proposed algorithm improves on prior work on Energy-based Fair Queuing (EFQ) \cite{b14,b15} and explores its potential to play a pivotal role in maximizing the user experience in battery-limited mobile systems. In prior work, by extending the traditional fair queuing to the energy domain, the EFQ scheduling algorithm was proposed to achieve proportional power sharing, effective time-constraint compliance, and a flexible trade-off between them. The behavior of the proposed algorithm is assessed on a test-bench based on different types of load. Based on the new test-bench, the properties of EFQ have been assessed in a more comprehensive way. In addition, EFQ has been compared with the default Linux scheduler of the Linux kernel V4.0.1 to show its advantage in maximizing the user experience in energy-limited mobile systems.

\section{Related Work}
In this section, the previous work most related to this paper is briefly introduced.

\subsection{Lifetime-Oriented Power Management}
Several lifetime-oriented Power Management (PM) schemes \cite{b6,b7,b8,b9} have been proposed to guarantee a target lifetime for battery-limited devices. Different from performance-centric PM schemes \cite{b2,b3,b4,b5}, these PM schemes are energy-centric in the sense that energy, instead of CPU or transmission bandwidth, is managed by the OS as the first-class resource in mobile systems. Neugebauer \textit{et al.} \cite{b7} and Flinn \textit{et al.} \cite{b9} demonstrated that a target lifetime can be achieved if the applications can self-adapt their energy demands based on the residual energy in the battery. However, the space to set a user-desired lifetime is limited by the degradation level of the applications. Moreover, the requirement of applications to be adaptive and energy-aware impedes these PM schemes from being widely applied in general systems. It has also been advocated that the battery discharge rate should be specifically restricted to reach the lifetime goal \cite{b6,b8}. Zeng \textit{et al.} \cite{b6} achieved the target lifetime by dividing it into a smaller number of time intervals named \emph{epochs} and limiting the energy available in each epoch; the limited energy in each epoch is allocated to different applications based on their energy demands. Roy \textit{et al.} \cite{b8} proposed the abstractions of \emph{reserve} and \emph{tap} to manage the energy on the basis of the applications: while \emph{reserve} stores an amount of energy for a target application, \emph{tap} places a rate limit on its energy consumption for guaranteeing the target lifetime. Although a wider range of battery lifetime is achievable by globally and abstractly managing energy as the first-class resource, little or no effort has been made to design an energy-based scheduler that is able to provide better support for time-sensitive tasks.

\subsection{Fair Queuing Scheduling}
Fair queuing has been widely applied in network scheduling \cite{b14,b15,b16} and CPU scheduling \cite{b17,b18,b19}. In the network domain, a fair queuing scheduler ensures that different packet flows can fairly share the output link to transmit data with a guaranteed rate. In the CPU scheduling domain, fair queuing algorithms are widely explored to support multimedia and other soft real-time applications on general-purpose operating systems. However, because fair queuing scheduling has no awareness of the time constraint, the performance of time-sensitive tasks is not guaranteed \cite{b18}. Several algorithms \cite{b18,b19} based on fair queuing have been developed to better support time-sensitive tasks by combining some extra real-time friendly mechanisms.

\subsection{Energy-Based Fair Queuing}
Energy-based fair queuing achieves proportional power sharing among schedulable entities by managing energy as the first-class resource and scheduling tasks based on their energy consumption. EFQ is an implementation of the Generalized Processor Sharing (GPS) \cite{b20} in the energy domain. The GPS is a classical fluid model that assumes system resources can be simultaneously served to multiple competing entities in infinitely divisible units. However, in a real system, the energy is served to tasks along with discrete time quanta whose length is constrained by the context switching overheads. The quantization of service time and energy brings inevitable energy allocation error, which is defined as the difference between the actual energy a task is served in the real system and the expected energy it should be served in the GPS model. Besides, a single device can be accessed by only one task at the same time, and the energy allocation of one task can be spent on diverse devices.

In this paper, the energy is correlated to the service time quantum by introducing the concept of an \emph{energy packet}. The energy packet $e_i^k$ is defined as the energy consumption of task $T_i$ during its $k$-th time quantum $q_i^k$. Then, the core of the EFQ implementation is to compute the \emph{starting energy tag} $S_i^k$ and the \emph{finishing energy tag} $F_i^k$, which separately trace the normalized energy consumption of task $T_i$ with weight $w_i$ before and after it is served the energy packet $e_i^k$. A Starting-Energy-based Fair Queuing (SEFQ) scheduler simply selects the task with the lowest starting energy tag.

Moreover, the SEFQ algorithm is enhanced from two aspects to provide better support for time-sensitive tasks. First, a \emph{share protection} mechanism is implemented on top of SEFQ to ensure a stable power share that is independent of the active task numbers. For any task that has a specific requirement on a stable power share $f_i$, it is reserved the desired power share by fixing the effective weight at $f_i$, while the effective weights of other tasks are recalculated based on their initially assigned weights (or simply initial weight) and the unreserved power share. Second, a real-time friendly mechanism named \emph{warp} is combined into SEFQ to give time-sensitive tasks priority on consuming their energy shares, and this new algorithm is called Borrowed Starting-Energy-based Fair Queuing (BSEFQ). To implement BSEFQ, two parameters, namely the \emph{warp value} and \emph{warp time limit}, are introduced into SEFQ. While a time-sensitive task is prioritized (or \emph{warped}) in consuming its energy allocation by subtracting the warp value from the starting energy tag $S_i^k$, the maximum length of time that the task can continuously have the priority is restricted by the warp time limit. In other words, BSEFQ integrates priority scheduling and fair queuing scheduling. It breaks short-term fairness to ensure strict time-constraint meeting for time-sensitive tasks; however, long-term proportional energy sharing among tasks is still maintained because the updating rate of the actual starting energy tag is restricted by the assigned weights. When there are multiple time-sensitive tasks in the system, different levels of warp value are determined for each task based on user preference or deadline-related parameters.

\section{The Proposed EFQ Algorithm}
The Completely Fair Scheduler (CFS) is the default scheduler for non-real-time task scheduling in Linux. Since CFS is a variant of fair queuing and shares the same principles with EFQ, its organizational structure has been maximally utilized to ease the code modification. The EFQ implementation work has been carried out in four steps.

\subsection{Preparation}
First, a new scheduling policy called \texttt{SCHED\_EFQ} is defined and correlated to the CFS structure. Then, EFQ-related variables, such as initial weight, reserved share, energy packet size, and warp parameters, are added into the scheduling entity structure \texttt{sched\_entity} and the run queue structure \texttt{cfs\_rq}; some energy measurement-related variables are also added to trace the total energy consumption as well as the individual energy consumption of each task. After that, the Linux niceness table is modified to be an array of 100 elements, which allows an extension of the nice levels to a range of $-50$ to $49$. The nice levels are mapped to the array index with a difference of 50; correspondingly, the kernel load weight stored in the niceness table is set within the range of 100 to 10{,}000 and increases in a step of 100, so that decreasing the nice level by 1 will increase the kernel load weight by 100. Since CFS schedules tasks based on the kernel load weight, the new niceness table allows the power share to be reserved in a resolution of 0.01 from the user space. In addition, it enables a quick matching between the EFQ effective weight and the Linux kernel load weight. Note that the default Linux niceness table only contains 40 values of kernel load weight, and they are mapped to the nice levels that range from $-20$ to $19$; the nice levels are further mapped to the global static priority by adding a difference of 120. As a result, the EFQ implementation also requires modifying the nice-to-priority translation with a new difference of 150. The above modifications are mainly taken place in the header files of the Linux scheduling sub-module.

\begin{figure}[!t]
\centering
\includegraphics[width=3.2in]{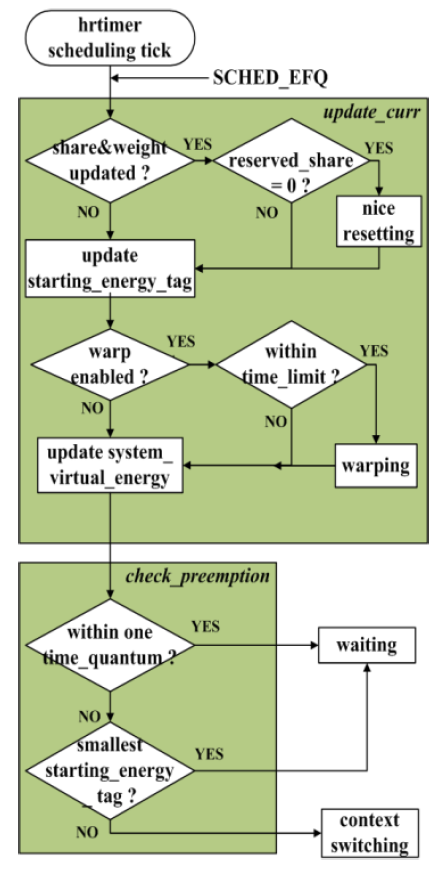}
\caption{Periodic scheduling tick.}
\label{fig:scheduling_tick}
\end{figure}

\subsection{Computation of Energy Tag and System Virtual Time}
\begin{figure}[!t]
\centering
\includegraphics[width=3.2in]{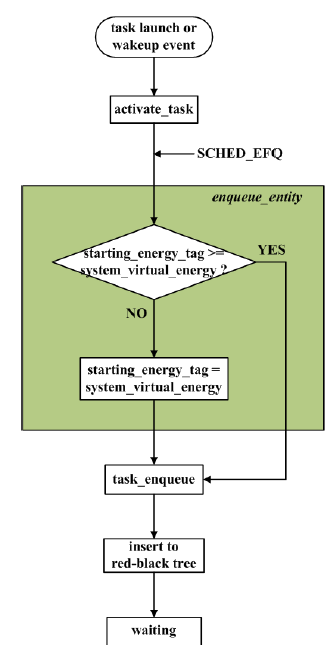}
\caption{Task launch or wake-up event.}
\label{fig:task_launch}
\end{figure}

This part of the work mainly takes place in the file \texttt{fair.c} under the Linux kernel directory \texttt{/kernel/sched}. Fig.~\ref{fig:scheduling_tick} shows the flow chart under the periodic scheduling tick, and Fig.~\ref{fig:task_launch} shows the flow chart upon a task join or wakeup.

In Fig.~\ref{fig:scheduling_tick}, the operations are periodically performed based on the scheduling tick that is generated by the Linux high resolution timer (\texttt{hrtimer}). Upon each scheduling tick, the \texttt{update\_curr} module in \texttt{fair.c} is called to update the execution time and EFQ-related variables of the current task if its scheduling policy is \texttt{SCHED\_EFQ}. Inside the function, the first assignment is to check the scheduling environment. For a task whose reserved share is zero, if there has been a change of the active task set, or the reserved share and initial weight of a certain task have been changed, the nice level as well as the kernel load weight of the task should be recomputed before updating the starting energy tag. However, for a task with a non-zero reserved share, its kernel load weight should remain the same, so no nice resetting is required. Next, the starting energy tag is updated based on the execution time, the energy packet size, and the kernel load weight of the current task. After that, a predefined warp value is subtracted from the starting energy tag if the task is time-sensitive and the continuous time it has been running with warp in the current period is less than the warp time limit. A time-sensitive task that reaches its warp time limit will keep the original starting energy tag; however, the task will be allowed to warp the energy tag another time once its next period starts. At the end of the \texttt{update\_curr} module, the system virtual energy is updated to the starting energy tag of the current task only in case that the task is not warped, because the system virtual energy has to trace the lowest starting energy tag and it should be monotonically non-decreasing.

Once the above operations have been done, the scheduler will check if the current task should be pre-empted by another task in the \texttt{check\_preemption} module. If the task execution time is within one scheduling time quantum, nothing is to be done until the next scheduling tick occurs; otherwise, extra functions are called to determine the next task to be dispatched by comparing the starting energy tag of the current task with the starting energy tag of the first task in the waiting queue \texttt{cfs\_rq}. If the current task has a smaller starting energy tag (especially when the task has a large power share or it is running with a warp), it will stay on the CPU for another time quantum; otherwise, the current task is forcibly removed from the CPU and inserted back to the waiting queue based on its starting energy tag, while the scheduler will make a context switch to the first task in the waiting queue. Note that the waiting queue in Linux is implemented through a red-black tree; a task with a smaller starting energy tag is inserted to the more-left position of the red-black tree, so that the leftmost task is the one with the smallest starting tag and is the first task in the waiting queue.

In Fig.~\ref{fig:task_launch}, an event of task-launch or task-wakeup will interrupt the current work in the system, and then the newly launched or woken-up task is activated. If the scheduling policy of the task is set as \texttt{SCHED\_EFQ}, the \texttt{enqueue\_entity} module in \texttt{fair.c} is called to update its starting energy tag. Because the starting tag of a newly-launched task is initialized as zero and the starting energy tag of a wakeup task does not increase while it is sleeping, the starting energy tag is set as the maximum value between the starting energy tag of the task and the system virtual energy. This prevents the newly-launched or wakeup task from continuously occupying the CPU with a considerably smaller starting energy tag. After the starting energy tag is updated, the activated task is inserted into the red-black tree of the associated waiting queue.

\subsection{Share Protection}
\begin{figure}[!t]
\centering
\includegraphics[width=3.4in]{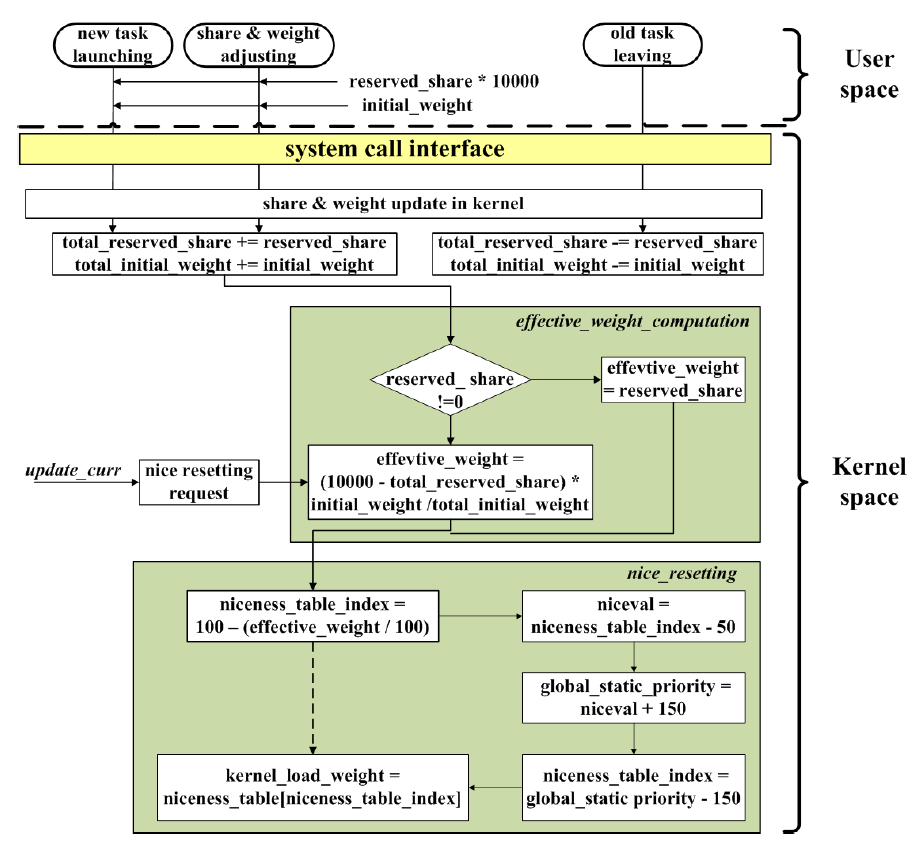}
\caption{Flowchart of share protection.}
\label{fig:share_protection}
\end{figure}

The code related to power share protection is added in the files \texttt{fair.c} and \texttt{core.c}. Fig.~\ref{fig:share_protection} demonstrates the flowchart of share protection. This part of the code is called upon four events: new task launching, share or weight adjusting, old task leaving, and the request of nice resetting from the \texttt{update\_curr} module. When a new task is launched, or the reserved share and initial weight of an old task are modified in the user space, the values of share and weight will be passed into the kernel space to update the corresponding variables. This update event will be later detected in the \texttt{update\_curr} module (Fig.~\ref{fig:scheduling_tick}) that is periodically performed based on the scheduler tick. To avoid incurring float computation in the kernel and reduce the scheduling overhead, the reserved share is scaled up by 10{,}000. After that, the \texttt{effective\_weight\_computation} module is called to compute the effective weight for the task. If the task is an important or time-sensitive one that is holding a non-zero reserved share, the effective weight is equal to the reserved share; otherwise, the effective weight is computed based on the total reserved share, the total weight of all active tasks, and the initial weight of the task. Finally, the effective weight is passed to the \texttt{nice\_resetting} module to compute the nice value and niceness table index, and the kernel load weight is set to be the indexed value (ranges from 100 to 10{,}000) of the niceness table. The above two modules will be called also from the \texttt{update\_curr} module when the nice value and kernel load weight of an active task should be reset. On the other hand, if an old task leaves the system after finishing its work, only the share and weight are updated in the kernel.

\subsection{System Calls}
\begin{figure}[!t]
\centering
\includegraphics[width=3.4in]{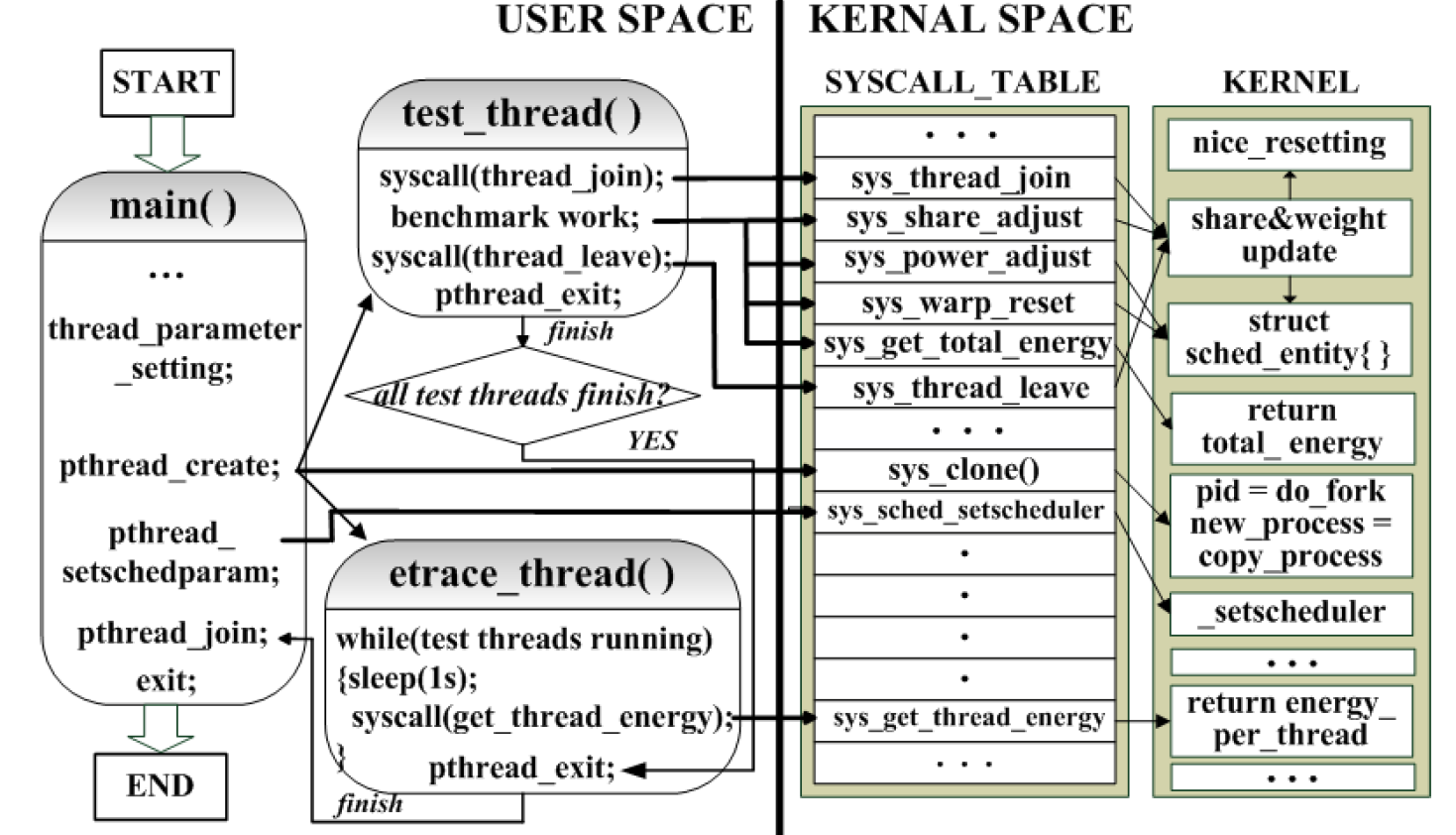}
\caption{Using system-call sequence.}
\label{fig:syscall_seq}
\end{figure}

The final step of the EFQ implementation concerns the modification of the system call interface (Fig.~\ref{fig:syscall_seq}). First, the existing system call \texttt{sys\_clone} and its related kernel functions are modified to allow the initialization of EFQ-related variables when a thread is created or a process is forked. Then, another existing system call \texttt{sys\_sched\_setscheduler} and its related function \texttt{sched\_setscheduler} are modified so that the new scheduling policy \texttt{SCHED\_EFQ} can be recognized and set as the policy of the created task structure; this system call is also used to pass EFQ parameters, such as warp, warp time limit, and energy packet size, to the task structure of a thread right after it is created. Besides, several new system calls have been implemented to enable the user-space thread to interact with its kernel task structure during the whole life cycle of the thread. Specifically, the system calls \texttt{sys\_thread\_join} and \texttt{sys\_thread\_leave} allow updating the share and weight in the kernel space when a new thread joins the system and an old thread finishes the work, respectively; the system call \texttt{sys\_weight\_adjust} allows the thread to adjust its share and weight in the middle of the thread execution; the system call \texttt{sys\_warp\_reset} allows resetting the warp value and warp time limit; and the system call \texttt{sys\_power\_adjust} allows changing the energy packet size of the thread based on the executed codes in the program.

\section{Experiments and Results}

\subsection{Environmental Setup}
The proposed algorithm was implemented as a replacement part of the Linux kernel scheduler code. Also, since the proposed algorithm is for battery-limited devices, all the experiments were carried out on laptop devices.

\subsubsection{Software Requirements}
\begin{itemize}
    \item \textbf{Operating System:} Any Linux distribution with kernel v4.0.1
    \item \textbf{Tool Required:} \texttt{powerstat} tool to measure power consumption by each process
\end{itemize}

\subsubsection{Hardware Requirements}
\begin{itemize}
    \item \textbf{Battery-Powered Devices:} The proposed algorithm was tested on three different laptop devices, namely Acer, SONY VAIO, and Dell systems.
    \item \textbf{Physical Memory:} 512 MB
\end{itemize}

\subsection{Experiment}
To test the performance of the proposed algorithm, the device was tested under different types of loads:
\begin{itemize}
    \item \textbf{No Load:} only background processes running.
    \item \textbf{Light Load:} one user-interactive process running (e.g., one browser tab with a website loaded).
    \item \textbf{Medium Load:} multiple user-interactive processes running (e.g., one browser with multiple tabs open).
    \item \textbf{Heavy Load:} various types of user-interactive processes (e.g., multiple browsers along with a Notepad file, Word document, Excel document, etc.).
\end{itemize}

\subsection{Performance Analysis}
As the power consumption of battery-limited devices depends on the device on which a process is running, the performance of the EFQ algorithm was tested on different types of battery-limited devices, including an Acer laptop, a Dell laptop, and a SONY VAIO laptop.

\begin{table}[!t]
\caption{Power Consumption (in Watts) on Acer Laptop}
\label{tab:acer}
\centering
\begin{tabular}{|l|c|c|}
\hline
\textbf{Load Type} & \textbf{Without EFQ (W)} & \textbf{With EFQ (W)} \\
\hline
No Load     & 16.41 & 14.17 \\
\hline
Low Load    & 21.40 & 19.55 \\
\hline
Medium Load & 29.47 & 24.39 \\
\hline
Heavy Load  & 32.25 & 39.13 \\
\hline
\end{tabular}
\end{table}

\begin{figure}[!t]
\centering
\caption{Comparison of EFQ with CFS on Acer Laptop.}
\label{fig:acer}
\end{figure}

Table~\ref{tab:acer} and Fig.~\ref{fig:acer} present the comparison of power consumption with and without the EFQ scheduler on the Acer laptop. It can be observed that for no-load, low-load, and medium-load scenarios, the EFQ algorithm consumes less power than the default CFS scheduler. However, under heavy load conditions, the CFS scheduler outperforms the EFQ algorithm in terms of power consumption.

\begin{table}[!t]
\caption{Power Consumption (in Watts) on Dell Laptop}
\label{tab:dell}
\centering
\begin{tabular}{|l|c|c|}
\hline
\textbf{Load Type} & \textbf{Without EFQ (W)} & \textbf{With EFQ (W)} \\
\hline
No Load     & 20.81 & 17.41 \\
\hline
Low Load    & 26.13 & 23.91 \\
\hline
Medium Load & 31.25 & 26.47 \\
\hline
Heavy Load  & 39.37 & 42.21 \\
\hline
\end{tabular}
\end{table}

\begin{figure}[!t]
\centering
\caption{Comparison of EFQ with CFS on Dell Laptop.}
\label{fig:dell}
\end{figure}

When compared to the Acer laptop's performance, the Dell laptop consumes more power for the same set of processes and type of load, as reported in Table~\ref{tab:dell} and Fig.~\ref{fig:dell}. However, like the Acer laptop, at heavy loads, the EFQ algorithm does not perform as expected, and the CFS performs better than EFQ.

\begin{table}[!t]
\caption{Power Consumption (in Watts) on SONY VAIO Laptop}
\label{tab:vaio}
\centering
\begin{tabular}{|l|c|c|}
\hline
\textbf{Load Type} & \textbf{Without EFQ (W)} & \textbf{With EFQ (W)} \\
\hline
No Load     & 14.27 & 11.91 \\
\hline
Low Load    & 19.57 & 16.17 \\
\hline
Medium Load & 26.17 & 23.71 \\
\hline
Heavy Load  & 32.87 & 35.12 \\
\hline
\end{tabular}
\end{table}

\begin{figure}[!t]
\centering
\caption{Comparison of EFQ with CFS on SONY VAIO Laptop.}
\label{fig:vaio}
\end{figure}

The results obtained on the SONY VAIO laptop are presented in Table~\ref{tab:vaio} and Fig.~\ref{fig:vaio}. The SONY VAIO laptop consumes the highest power among the three when compared to the Acer and Dell laptops for the same set of processes and type of load. However, similar to the other two systems, at heavy load conditions, the EFQ algorithm does not perform as expected, and the CFS performs better than EFQ. Overall, the SONY VAIO system consumes the highest power when compared to the other two laptops.

From the experiments, it is evident that EFQ provides effective proportional power sharing under no-load, light-load, and medium-load conditions across all three battery-limited devices. The slight degradation under heavy load is attributable to the increased scheduling overhead, the energy estimation error, and the higher contention for energy among a larger number of active tasks.

\section{Conclusion}
The Energy-based Fair Queuing (EFQ) algorithm has been implemented based on Linux, and a test-bench based on various types of load has been utilized to assess the EFQ algorithm. As verified by the experimental results, EFQ can achieve proportional power sharing by accounting for energy consumption on both CPU and I/O operations and scheduling tasks based on their normalized energy consumptions. Also, it has been demonstrated that EFQ can protect the power share of certain specific applications, which can enable the achievement of advanced energy-related objectives that are impossible with traditional processor schedulers like CFS. Time-constraint compliance under EFQ depends not only on the amount of energy allocation, but also on how timely the energy is scheduled to time-sensitive tasks. Experimental results show that SEFQ can achieve good real-time performance when the energy estimation is accurate and the number of energy-competing tasks is small. However, the time-constraint compliance under SEFQ has weak robustness when the active task number and energy estimation error increase. This problem can be solved in BSEFQ without overly reserving power shares for time-sensitive tasks. In the future, energy estimation and accounting will be implemented to feed the energy information in real-time to the EFQ scheduling; and energy consumption on more devices should be explored to assess the system-wide energy management of EFQ algorithms.

\bibliographystyle{IEEEtran}
\bibliography{ref}

\end{document}